\documentclass[mamp,a4paper]{jpconf}
\usepackage{graphicx}
\usepackage{epsfig}
\usepackage{bm}

\begin{document}
\title{Microquasar LS 5039:  a TeV gamma-ray emitter 
and a  potential TeV neutrino source\footnote{The material of this paper was one of the highlight topics of 
the Plenary talk of F.A. Aharonian on the theory of TeV gamma-ray sources 
at the TAUP05 Conference. Instead of a brief overview of models of all 
classes of detected gamma-ray sources, in this article we decided to focus 
on high energy processes in the microquasar LS5039 which presents a great 
interest for the Astroparticle Physics community in general, and for 
gamma- and neutrino astronomers, in particular.}}
\author{F Aharonian$^1$, L Anchordoqui$^2$, D Khangulyan$^1$ and T Montaruli${}^{3,4}$}
\address{$^1$Max-Planck-institut f\"ur Kernphysik, Saupfercheckweg 1, 69117 Heidelberg, Germany}
\address{$^2$Department of Physics,
Northeastern University, Boston, MA 02115}
\address{$^3$Department of Physics,
University of Wisconsin, Madison WI 53706}
\address{$^4$Universit\'a di Bari, Via Amendola 173, 70126 Bari, Italy}

\ead{felix.aharonian@mpi-hd.mpg.de}


\begin{abstract}
\noindent 
The recent detection  of TeV $\gamma$-rays  from 
the microquasar LS 5039 by \textit{HESS}  
is one of the most  exciting discoveries  of observational  
gamma-ray astronomy  in the very high energy regime.  
This result  clearly demonstrates  that $X$-ray  binaries with 
relativistic jets  (microquasars) are 
sites of  effective acceleration of  particles (electrons and/or protons) 
to multi-TeV energies.  Whether the $\gamma$-rays are of hadronic 
or leptonic  origin is a key issue related to the origin of 
Galactic Cosmic Rays.
 We discuss different possible scenarios for 
the  production of $\gamma$-rays, and argue in favor of 
hadronic origin  of TeV photons, especially if they are 
produced within the binary system. If so, 
the detected $\gamma$-rays should be accompanied 
by a flux of high energy neutrinos emerging 
from the decays of $\pi^\pm$ mesons produced at $pp$ and/or
$p \gamma$ interactions. The flux of TeV 
neutrinos, which  can be estimated on the basis of the detected 
TeV $\gamma$-ray flux, taking into account the 
internal $\gamma \gamma \rightarrow e^+e^-$ absorption, 
depends significantly on  the location of $\gamma$-ray production
region(s).  The minimum neutrino flux  above 1 TeV is expected to be 
at the level of $10^{-12} \ \ \rm cm^{-2} s^{-1}$; however,
it could be up to a factor of 100 larger.  The detectability of the signal 
of multi-TeV neutrinos significantly depends on the high energy cutoff
in the spectrum of parent protons; if  the spectrum of accelerated 
protons continues to 1 PeV and beyond,  the predicted neutrino 
fluxes can  be probed by the planned km$^3$-scale neutrino 
detector. 
\end{abstract}

\section{Introduction}
During the next decade, two  km$^3$  scale underwater/ice 
neutrino telescopes will  start operation, both in the Northern 
(NEMO/km3NeT~\cite{Piattelli:2005mf}) and Southern (IceCube~\cite{Ahrens:2003ix}) 
hemispheres.  Theoretical and phenomenological studies of 
recent years show that the sensitivities of these detectors
should  allow meaningful probes of different
nonthermal galactic and extragalactic source populations
in high energy neutrinos (see e.g.~\cite{Halz_Hooper,Bednarek:2004ky}).

TeV neutrinos are unique
carriers of unambiguous information about hadronic processes 
in cosmic accelerators. They  are produced in 
$pp$ or $p \gamma$ interactions through the decay of secondary 
charged pions.
Since generally these processes also result 
in TeV $\gamma$-rays of a comparable rate, 
the best candidates to be discovered  by planned $\rm km^3$ 
scale neutrino 
telescopes are  sources of $\gamma$-rays with fluxes
above 1 TeV at the level of 
$(3-10) \times 10^{-12}~{\rm ph}/ \, {\rm cm}^{2} {\rm s}$.  

Presently more than three  dozens of  sources of both 
galactic and  extragalactic origin are established  as 
TeV $\gamma$-ray emitters.  
Some of them, in particular  four galactic
objects -  RXJ1713.7-3946 and RXJ0852.0-4622 ( shell type  
supernova remnants)
and Crab Nebula and Vela X (plerions), 
as well as three  variable extragalactic (BL Lac) objects  -   
Mrk 421, Mrk 501,  and  1ES1959+650, do exhibit persistent 
or episodic fluxes of TeV $\gamma$-rays exceeding
$J_\gamma(E\geq 1 \rm TeV )=
10^{-11} \  \rm   ph \, cm^{-2}\, s^{-1}$.  
Thus, if a significant fraction of TeV $\gamma$-ray 
fluxes of these sources are of hadronic origin, and the spectra of 
parent protons extend to  PeV energies,  
IceCube and  NEMO/km3Net  should be able to  detect 
the neutrino counterparts of TeV $\gamma$-rays. 

Moreover, unlike neutrinos,  
TeV $\gamma$-rays are fragile and suffer 
strong internal and/or  external
absorption due to interactions with
ambient  IR/optical photon fields.     
Therefore, for sources with heavily 
absorbed TeV emission,  
the chances of detection of counterpart neutrinos 
could be quite high, even for relatively 
weak $\gamma$-ray sources  with a TeV flux 
below  $10^{-12}~{\rm ph}/{\rm cm}^{2}\, {\rm s}$. 
In this regard, the microquasar 
LS~5039~\cite{Paredes2000} recently detected 
in TeV $\gamma$-rays by the 
\textit{HESS} Collaboration~\cite{Aharonian2005} 
is of great interest and a promising source  
also for neutrino astronomy. 

\section{High Energy Nonthermal Phenomena in X-ray Binaries}

In one of the first attempts to  classify the potential VHE $\gamma$-ray emitters,  
X-ray binaries  were attributed  to the category of serendipitous sources, 
i.e. objects  which ``do not have any firm   \textit{a priori}
basis for their selection as candidate 
VHE $\gamma$-ray sources'' \cite{Weekes92}.
Nevertheless, this very class of objects played a very  important  role
in gamma-ray astronomy,   in particular initiating  in the 1980s a  renewed interest 
in  ground based  $\gamma$-ray  observations.  
But ironically, the same sources ultimately raised questions about the 
credibility of the  results  from that controversial  period of 
ground-based $\gamma$-ray astronomy. 
As a result,  since the early 1990s  X-ray binaries have not been 
considered  as primary targets for VHE $\gamma$-ray  observations.  
This pessimistic view was largely  supported also   
by the belief that   X-ray binaries should be treated, first of all,   
as {\em thermal}  sources effectively  transforming  the gravitational 
energy of the compact object  (a neutron star or  a black hole) into 
thermal X-ray emission radiated  away  by the hot accretion plasma.  

After  the discovery of galactic sources with 
relativistic jets  dubbed  {\em Microquasars}~\cite{Mirabel_Rodriguez_1994},
the general view  on  X-ray binaries 
has  dramatically  changed;  the observations of synchrotron 
radio flares from microquasars demonstrated non-negligible role 
of particle acceleration (MeV/GeV electrons) in these 
accretion-driven objects.  The discovery of  
 $\gamma$-rays  from LS~5039  provides a clear indication that nonthermal 
phenomena of particle acceleration and radiation  
in  X-ray binaries  effectively extend into the TeV and, perhaps,  also PeV 
energy regimes. In this regard,  many original ideas and models 
(see e.g. \cite{Berezinsky,Eichler,Hillas,Mitra,BeamCloud}),  which were 
initiated by the early  (wrong ?) claims of TeV and PeV signals from Cyg X-3 and Her X-1
(for a review see  \cite{Weekes92}),   remain very attractive and applicable, after some 
modifications,  to  Microquasars as well.

Variable TeV  radiation  has been predicted from microquasars  in
different astrophysical scenarios with involvement of both leptonic~\cite{LeptonicGamma1} 
and hadronic~\cite{HadronicGamma1} interactions.
Gamma-rays of leptonic~\cite{LeptonicGamma2} and hadronic \cite{Heinz,Bosch2005a}
origin associated with termination of jets in the interstellar medium 
are expected also from extended regions surrounding microquasars. 

Hadronic models are attractive in
the sense that they predict,  from $X$-ray binaries in general, and from
microquasars in particular, significant  fluxes of both TeV 
$\gamma$-rays and  neutrinos due to proton-proton,  photomeson and
photo-disintegration processes that may take place in the wind and/or the 
atmosphere of the normal star, in the accretion
disk, and in the jet (see e.g.~\cite{Berezinsky,Eichler,Hillas,Mitra, 
BeamCloud,Neutrinos2, Neutrinos3, Neutrinos4, Neutrinos5, Neutrinos6, Neutrinos7}). 

While the models of production of TeV neutrinos 
demonstrate the feasibility of several different scenarios of particle 
acceleration and  interactions in microquasars, one should not 
overestimate   the  prediction power  and robustness of  calculations of 
neutrino fluxes based merely  on  model 
assumptions such as (i) the total nonthermal energy  of the source, (ii) the
efficiency of the particle acceleration, (iii) the energy spectrum and the 
maximum energy of accelerated protons, 
(iv) the efficiency of conversion of proton energy into neutrinos.  
Obviously, the flux  predictions  based on many model assumptions and 
parameters  does not guarantee any reliable estimate of neutrino fluxes. 
While  this concerns (perhaps, with some exceptions)  all astrophysical 
source populations, in the case of microquasars 
several   model assumptions together result in flux 
uncertainties as large as orders of magnitude.  
Moreover,  in microquasars  particles  can be  boosted   to
multi-TeV energies only if the acceleration proceeds at the theoretically highest
possible rates allowed by classical electrodynamics.   It is not obvious, 
however,  that this should  work, from first principles,   in microquasars. 

The  detection of  TeV $\gamma$-ray emission   from LS~5039  
has dramatically  changed the  status of model predictions. 
First of all,  the \textit{HESS} discovery provides the first unambiguous 
evidence for presence of multi-TeV  particles in microquasars. Below we 
demonstrate that if gamma-radiation is produced within the binary
system, then (2) electrons hardly can be accelerated to multi-TeV energies
to explain the observed TeV $\gamma$-ray signal, and, therefore,
(3) the parent particles of $\gamma$ -rays should be protons or nuclei. 
If so, (4) $\gamma$-radiation should be accompanied by TeV neutrinos with a 
production rate comparable to the gamma-ray production rate. 
Furthermore, since the $\gamma$-rays
suffer unavoidable absorption or cascading,
 (5) the flux of neutrinos should significantly exceed
the observed flux of $\gamma$-rays, and  therefore (6) can be (marginally) 
detectable  by the km$^3$ class high energy neutrino telescopes.

\section{The case of LS~5039}

While the \textit{HESS} discovery
of TeV $\gamma$-ray emission from LS 5039 does support the early theoretical
speculations of acceleration of particles in microquasars to TeV/PeV  energies,
the reported data do not allow us to relate the $\gamma$-ray production
region(s) to specific sites of this complex system. The upper limit on the
source angular size of about 50 arcsec~\cite{Aharonian2005} 
implies that, for a source distance $d
\simeq 2.5$~kpc~\cite{Casares2005}, $\gamma$-ray production takes place
within a radius $\sim 0.6$~pc around LS~5039.  Although the observed flux
does not allow an unequivocal conclusion about the variability of the TeV
source~\cite{Aharonian2005}, hints have been recently
reported~\cite{Casares2005} of a possible correlation of the TeV flux with 
the $3.9$ day orbital period of the binary system. If confirmed
by future observations, this would be an indication that 
TeV $\gamma$-rays are produced in a compact region, presumably 
inside the binary system.  In what follows we will assume that 
$\gamma$-rays are produced within $R \sim 10^{13} \ \rm cm$, 
which automatically implies very strong $\gamma \gamma$ absorption. The
effect of this absorption should have a strong impact on the detectability of
TeV neutrinos from this source, provided that detected $\gamma$-rays are of
hadronic origin.

TeV  $\gamma$-ray  production generally 
requires two components: $(a)$ an \textit{effective accelerator} 
of particles, electrons and/or protons,  
up to 10 TeV and beyond; $(b)$ an \textit{effective target}
(converter). 
While the most likely site 
for particle acceleration in LS~5039 is the 
jet, which  with a speed $v = 0.2 \ c$
and a  half-opening angle
$\theta \leq 6^\circ$ extends out to 300 mas
($\approx 10^{16}$ cm)~\cite{Paredes2002},
the bulk of $\gamma$-rays could  be produced 
both inside and outside the jet. 
In particular,
relativistic electrons in the jet can be accelerated through internal 
shocks~\cite{Kaiser2000}, and consequently high energy $\gamma$-rays 
can be effectively produced through inverse 
Compton scattering~\cite{Levinson95,LeptonicGamma1,Paredes2000,Bosch2005b,Dermer_Bottcher05}. 
However, in the jet of LS~5039,  because of  close location of the very luminous companion star,
the effects of fast radiative cooling may  prevent the electrons 
from reaching multi-TeV energies.

The maximum energy of accelerated electrons is achieved 
when the acceleration time approaches the cooling time 
due to  radiative 
(synchrotron and Compton) losses.  The acceleration time 
can be presented in a general form
$t_{\rm acc}=\eta\, r_{\rm L}/c 
\approx 0.11\, E_{\rm TeV}\, B_{\rm G}^{-1}\, \eta \ \rm s$,
where $r_{\rm L}=E/eB$ is the Larmor radius,
$E_{\rm TeV}=E/1 \rm TeV$, 
and $B_{\rm G}=B/1 \rm G$ is the strength of the ambient 
magnetic field. The parameter $\eta$ characterizes 
the efficiency of acceleration: in the case of 
extreme accelerators (maximum possible acceleration rate
allowed by classical electrodynamics) $\eta \rightarrow 1,$ whereas for  
shock acceleration in the Bohm diffusion regime 
$\eta \approx 10 (v/c)^{-2}$ (see e.g.~\cite{Malkov:2001}). 
For a star with a temperature $kT=3.5$~eV and optical luminosity 
$L_\star \approx 1.2 \times 10^{39} \ \rm erg/s$, 
the energy density of  the starlight close to the compact object, 
$w_r=L_\star/4 \pi R^2 c$, varies between   $400$ and  
$1600 \ \rm erg/cm^3$  at the apastron 
($R \approx 2.9 \times 10^{12} \ \rm cm$), and at the periastron 
($R \approx 1.4 \times 10^{12} \ \rm cm$), 
respectively~\cite{Casares2005}.
The Compton scattering of TeV electrons in the field of 
$3 kT \approx 10$~eV starlight 
takes place in deep Klein-Nishina regime. While in the Thompson regime 
($E_e \ll 0.1$ TeV) the cooling time is inverse proportional to the electron energy, 
$t_{\rm T}\approx 0.03 E_{\rm TeV}^{-1} \ \rm s$, in the Klein-Nishina regime 
the characteristic Compton cooling time 
can be approximated with a good accuracy, 
$t_{\rm C} \approx 34  
w_0^{-1} E_{\rm TeV}^{0.7} \ \rm s$, 
where $w_0= w_{\rm r}/500~{\rm erg} 
\ {\rm cm}^{-3}$.   
From   $t_{\rm acc} = t_{\rm C}$ one finds 
\begin{equation}
E_{\rm e,max} \simeq  
2 \ [B_{\rm G} (v/0.2c)^2  
w_0^{-1}]^{3.3}~{\rm TeV}.
\label{Emax1}
\end{equation}
Formally,  for $B > 1$~G the maximum energy of
accelerated electrons can exceed 10 TeV.
However, for such a large $B$-field, the synchrotron losses
dominate over the Compton losses.
Namely, the condition  $t_{\rm acc} = t_{\rm sy}$, where 
$t_{\rm sy} \approx 400 B_G^{-2} E_{\rm TeV}^{-1} \ \rm s$
is the synchrotron  cooling time, gives 
\begin{equation} 
E_{\rm e,max}  \approx 3.8 \, B_{\rm G}^{-1/2} \, (v/0.2c)~{\rm TeV} \ .
\label{Emax2}
\end{equation}
Equations~(\ref{Emax1}) and (\ref{Emax2}) imply that 
electrons could be accelerated to energies exceeding 
10 TeV only in an environment with 
$B \leq 0.1$~G and $w_0 \ll 1$. 
Such small fields can be  realized  beyond the binary system,
in outer parts of the jet.  
In this case, $\gamma$-rays 
can be produced  through the Synchrotron-self-Compton (SSC)
scenario~\cite{LeptonicGamma1}, although the   
contribution from  the external (starlight)  
photons could be still significant even  at distances 
quite far from the  binary system. It was recently argued~\cite{Dermer_Bottcher05} 
that the inverse Compton on starlight cannot explain the hard TeV $\gamma$-ray 
spectrum, and, therefore,  an  additional  SSC component was proposed to match 
the \textit{HESS} data.  However,  the SSC model does not solve the major 
problem associated with the maximum achievable energy of electrons. In fact, 
the SSC model 
requires even higher electron energies than the external Compton model.

Thus, any evidence of $\gamma$-ray 
production inside the binary system, e.g. 
detection of a periodic component of TeV radiation 
would be a strong argument in favor of the hadronic 
origin of these energetic photons. The extension of hard $\gamma$-ray 
spectrum above several~TeV requires acceleration of protons 
to $\geq 100$~TeV. Within the inner parts of the jet,  
with a radius $R_{\rm jet} \sim 10^7-10^8$~cm, 
a magnetic field $B \geq 10^5~{\rm G}$
could be sufficient to boost 
protons up to very  high energies. 
The maximum  energy is determined by 
the condition $r_L \leq R_{\rm jet}$, which  gives 
$E_p \leq 3 \times 10^{15} 
(R_{\rm jet}/10^8~{\rm cm})(B/10^5~{\rm G})  \ \rm eV$.
Although these protons are not sufficiently energetic to 
interact with the starlight photons, they can 
trigger photomeson processes on $X$-ray photons in the accretion disk.
These interactions may lead to copious production
of high energy neutrinos, neutrons, $\gamma$-rays and electrons. While neutrons and 
neutrinos  escape the source without significant suppression,
the electromagnetic fraction of the energy is 
effectively reprocessed,  
and escape the source mainly  in the form of hard
$X$-rays and low energy $\gamma$-rays. 
For example, if the luminosity of the accretion disk
at UV band exceeds $10^{35} \ \rm erg \,s^{-1}$, 
all $\gamma$-rays above several tens of~GeV will be 
converted into $e^\pm$ pairs before they 
can leave the source. Hence, the synchrotron radiation of 
the last generation of electrons will lead to a hard 
$\gamma$-ray spectrum $\propto E^{-1.5},$ extending to 
$\sim 100$~MeV.  Interestingly, observations of LS~5039 
performed by the COMPTEL telescope do show~\cite{Collmar:2004}
a statistically significant signal of MeV $\gamma$-rays 
with a photon index $1.6\pm 0.2$ and a total energy flux between
1 MeV and 30 MeV of about $9 \times 10^{-10} \rm \ erg/cm^2 s$.
If this radiation is indeed  related to LS~5039, 
it would imply a pronounced maximum 
in the spectral energy distribution at the level of 
$6 \times 10^{35} \ \rm erg \ s^{-1}$. 

Thus, speculating  that this radiation is 
initiated by photomeson interactions (of protons with energy $\geq 100$ TeV) 
close to the inner part of the accretion disk,   
one can estimate the expected flux 
of TeV muon neutrinos at the level of 
$\sim 10^{-10} \ \rm cm^{-2} \, s^{-1}$, provided that 
the charged pions decay before interacting with the ambient 
dense plasma and radiation.  
In the lab frame, the decay time of  
charged pions responsible for TeV neutrino production is 
$t_{\pi^\pm}= (E_\pi/ m_\pi c^2)\,\tau_{\pi^\pm} 
\approx 2.5 \times 10^{-3} (E_\pi/10 \ \rm TeV)$~s.  
On the other hand, the cooling time  of $\pi^\pm$ 
due to inelastic $\pi p$ and $\pi \gamma$ 
interactions depends on the ambient gas $n_p$ 
and photon $n_x$ densities:  
$t_{\pi p} \sim 10^{-3} (n_p/10^{17}~{\rm cm^{-3}})^{-1}$ s, 
and 
$t_{\rm \pi \gamma} \sim 5 \times 10^{-3} 
(n_X/10^{20} {\rm cm^{-3}})^{-1}$ s.
For typical parameters characterizing LS~5039, 
the number density of $X$-ray  
photons and the plasma density
in the region  $R \geq 10^7$~cm do not exceed 
$10^{20} \ \rm cm^{-3}$ and 
$10^{17} \ \rm cm^{-3}$, respectively.   
Therefore charged pions decay to $\mu$ and $\nu_\mu$
before interacting with the ambient photons and protons. 
The production of neutrinos from the subsequent muon decay
also proceed with high probability as long as the magnetic 
field does not exceed $B \sim 10^6$~G. This follows 
directly from the 
comparison of the decay time of muons,  
$t_\mu= (E_\mu/m_\mu c^2)\tau_{\mu} \simeq 
0.2(E_\mu/10 \rm TeV)$~s, with their
synchrotron cooling time
$t_{\rm sy} \approx 0.07 
\,(B/10^6 \rm G)^{-2} (E_\mu/10~{\rm TeV})^{-1}$~s.

Protons can also interact effectively with the ambient cold plasma, close to the
base of the jet and/or throughout the entire jet. In what follows we assume that
the base of the jet is located close to the inner parts of the accretion disk,
i.e., the jet axis $z$ is taken normal to the orbital plane, with $z_0 \sim 30
R_{\rm S}$ ($R_{\rm S} \simeq 3 \times 10^5 (M_{\rm BH}/M_\odot)  \rm cm$ is the
Schwarzschild radius). If the magnetic field drops as $B \propto z^{-1}$,
the condition of the confinement of protons in the jet, $r_{\rm L} \leq   R$, 
where $R=\theta z$ is the radius of the jet at a distance $z$, implies
$E_{\rm max} \propto B z$=constant. Thus, one  may expect 
acceleration of protons  to the same maximum energy 
$E_{\rm max}$  over the
entire jet region. However, if there is a faster drop of $B$ with $z$, the
protons at some distance $z_t$ from the compact object will start escaping the
jet. If this happens within the binary system, i.e.  $z_t \leq 10^{12} \rm cm$,
protons interacting with the dense wind of the optical star will result in
additional $\gamma$-ray and neutrino production outside the jet.

If the jet power is dominated by the 
kinetic energy of bulk motion of cold $e$-$p$ plasma, 
the baryon density of the jet  
$n_{\rm jet}$ can be estimated from the condition 
$L_{\rm jet}=\pi\, R_{\rm jet}^2(z) 
\,n_{\rm jet}(z)\, m_p v^3/2$.
The efficiency of $\gamma$-ray 
production in the jet is  
$\rho_\gamma= L_\gamma/L_p = \sigma_{pp} f_\pi 
\int_{z_0}^{z_t} n_{\rm jet}(z) dz \leq 1$,
where $L_\gamma$ is the luminosity of VHE $\gamma$-rays 
and $L_p$ is the  power of accelerated protons. Here,  
$\sigma_{pp} \approx 40$~mb is 
the cross-section of inelastic  $pp$ interactions, 
and $f_\pi \approx 0.15$ is the fraction of the 
energy of the parent proton transfered to a 
high energy $\gamma$-ray photon. Given the recent estimate 
of the black hole mass in LS~5039 $M=3.7_{-1.0}^{+1.3} \, M_\odot$~\cite{Casares2005}, we set  
$z_0 \simeq 3 \times 10^7 \ \rm cm$.
For the profile of the number density  we adopt a power law-form, 
$n_{\rm jet} = n_0 (z/z_0)^{-s},$ where $s=0$ corresponds to a cylindrical geometry, $s=2$  to 
a conical jet, and $s \sim 1$ is an intermediate case. 
Expressing the acceleration power of protons 
in terms of the total jet power, 
$L_p=\kappa L_{\rm jet}$, 
one  finds the following requirement for the jet power,   
 \begin{equation} 
 L_{\rm jet} \approx 2 \times 10^{37}\,
\frac{L_{\gamma,34}^{1/2} (v/0.2c)^{3/2}}  
 {\sqrt{C(s) (\kappa/0.1)}}
 \ \rm erg \ s^{-1} \ ,
 \label{jetpower}
 \end{equation} 
where  $\kappa$ is the acceleration efficiency and
$L_{\gamma,34}= L_{\gamma}/10^{34} \ \rm erg \ s^{-1}$. 
The parameter $C(s)$   
characterizes the geometry/density profile 
of the jet. For $s=0, \ 1, \ 2$, one has
$C(s)=z_t/z_0, \, \ln(z_t/z_0),$ and 1, respectively. 
The case of
cylindrical jet provides the highest efficiency of
$\gamma$-ray production. However, since
$L_\gamma \leq 1/30 L_{\rm jet}$ (assuming $\approx 10\%$
efficiency of proton acceleration, and taking
into account that the fraction of energy of protons
converted to $\gamma$-rays cannot exceed 30\%)
the  $\gamma$-ray production cannot be extended
beyond $z_t \sim 10^4 z_0 \sim 3 \times 10^{11} \rm cm$.
The case of a conical jet  corresponds to the minimum
efficiency of $\gamma$-ray production, and thus the
largest kinetic power of the jet. In this case the bulk of
$\gamma$-rays are produced not far from the base.
Finally, in  the intermediate case,
TeV $\gamma$-rays are produced in equal amounts  per
decade of length of the jet, until
the jet terminates.

\begin{figure}
\centering\includegraphics[height=12cm,angle=270]{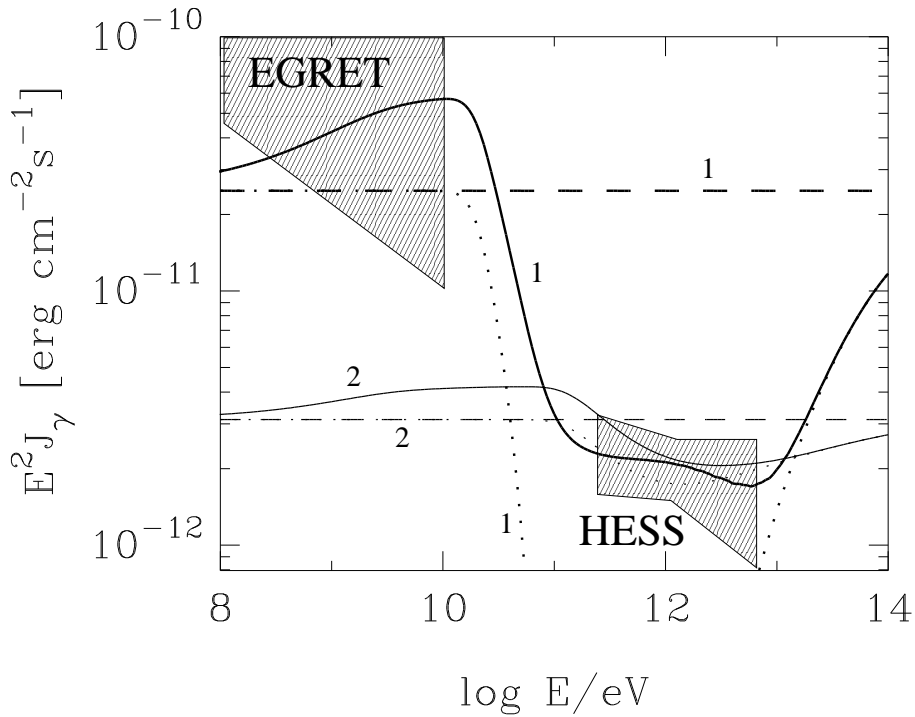}
\caption{Time averaged  
$\gamma$-ray spectra of LS~5039 
formed due to absorption (dotted lines) and cascading (solid lines)
in the anisotropic radiation field of the normal companion star. 
Curves 1 and 2 correspond two  
different assumptions on the location of the $\gamma$-ray production region  
in the jet: at $z=10^8$~cm (1), and $z=10^{13}$~cm 
(2). Dashed curves correspond to the primary $\gamma$-ray injection rates 
chosen in a way that the calculated cascade spectra 
match the observed $\gamma$-ray flux around 1 TeV.   
The shaded regions indicate the ranges of MeV/GeV (EGRET) and 
TeV (\textit{HESS}) $\gamma$-ray fluxes~\cite{Aharonian2005}.}
\label{fig}
\end{figure}

If $\gamma$-rays are
indeed produced in $pp$ interactions, one would expect production of high
energy neutrinos with a rate close to the $\gamma$-ray production rate, 
$Q_{\nu_\mu}(E)=\zeta Q_\gamma (E)$, where 
$\zeta$ varies between 0.5 and 2  
depending on the shape of the proton spectrum. However, 
since $\gamma$-rays are subject to energy-dependent absorption~\cite{Aharonian2005},
both the energy spectrum and the absolute flux of 
neutrinos, $J_{\nu_\mu}(E) \simeq \zeta J_\gamma(E)\, \exp[\tau(E)],$
could be quite different from the detected $\gamma$-rays, 
$J_\gamma(E) \approx 1.2 \times 10^{-12}
E_{\rm TeV}^{-2.1}~{\rm ph}\,{\rm cm}^{-2} 
\,{\rm s}^{-1}\, {\rm TeV}^{-1}$~\cite{Aharonian2005}.
The optical depth $\tau (E)$ depends
significantly on the location of 
the $\gamma$-ray production region, and
therefore varies with time if this region occupies a small volume of the
binary system.  This may lead to time modulation of the energy spectrum 
and the absolute flux of TeV radiation with the orbital period~\cite{Bottcher:2005pj,Dubus}. 
However, the $\gamma \gamma$ interactions generally
cannot be reduced  to a simple effect of absorption. In fact, 
these interactions initiate high energy electron-photon
cascades, supported  by the inverse Compton scattering
and $\gamma \gamma$ pair production processes. The cascades
significantly increase the transparency of the source.
The spectra of $\gamma$-rays formed during
the cascade development significantly differ from the
spectrum of $\gamma$-rays that suffer only absorption. 
This is shown in Fig.~\ref{fig}.  
Three processes have been included in the calculations - photon-photon
pair production, inverse Compton scattering and synchrotron radiation
of electrons.  The calculations of the electromagnetic cascade
developed in the photon and magnetic field are based on the method
similar to the one described in Ref \cite{3cascades}.  Because of the
orbital motion, both the absolute density and the angular distribution
of the thermal radiation of the star relative to the the position of
the compact object (microquasar) vary with time. In the calculations
we take into account the effect of the anisotropic (time-dependent)
distribution of target photons on the Compton scattering and
pair-production cross-sections \cite{Khangul_Ah05}.  In Fig.~\ref{fig}
we show the $\gamma$-ray spectra averaged over the orbital period.

The solid curves show the cascade spectra, the 
dotted curves are the spectra of $\gamma$-rays 
which the observer would see due to the pure absorption effect
(i.e. in the case of effective suppression of the cascade), and 
the dashed curves indicate the injection spectra. 
Two different locations of  $\gamma$-ray production region
have been assumed:  
(\textit{i}) close to the base of the jet  $z \leq  10^8$~cm, 
and (\textit{ii}) well above the base of the jet,  $z=10^{13}$~cm.   
All curves are averaged over the orbital period of the system,
taking into account the recent data
concerning the geometry of the system~\cite{Casares2005}.  
The cascade spectra are normalized to the 
reported range of TeV fluxes. They correspond
to the production rate (luminosity) of $0.1-10$~TeV $\gamma$-rays
between $10^{34}$ (corresponding to the dashed curve~2 for $d=2.5$ kpc) 
and $8 \times 10^{34}  \rm erg \ s^{-1}$ (dashed curve~1).
Figure~\ref{fig} shows that  
the cascade $\gamma$-rays calculated for $z=10^8$~cm agree quite  
reasonably with the low energy (MeV/GeV) data as well, 
given the large statistical uncertainties and 
the fact that the EGRET and \textit{HESS} observations are from different 
epochs. 

Since muon neutrinos should be produced with 
about the same rate as TeV $\gamma$-rays,
the dashed curves  can be used to estimate
the neutrino flux expected from the source:  
$J_{\nu_\mu} (>1 \rm  TeV) = 1.6  \times 10^{-11}$ 
and $1.9 \times 10^{-12} 
{\rm cm}^{-2} \ {\rm s}^{-1}$ for 
the cases (\textit{i}) and (\textit{ii}), respectively. 
The flux of neutrinos could be, in principle,  higher
if there is suppression of the cascade, as indicated by the dotted curves. 
This could happen if the magnetic field within the 
binary system exceeds 30~G. Although the energy 
density of this field $B^2/8 \pi \approx 35 \ \rm 
erg \ cm^{-3}$ is significantly less than the energy density
of the  radiation field, due to the Klein-Nishina 
effect, the cooling of the electrons is dominated by 
synchrotron radiation 
in the energy interval  $E \approx 100 \ {\rm keV} - 10 \ {\rm MeV}$. 
The flux observed by COMPTEL~\cite{Collmar:2004} provides an upper bound on the energy 
released via synchrotron radiation, and consequently sets an upper bound on the 
expected TeV neutrino flux around $J_{\nu_\mu} (> 1~{\rm TeV}) = 10^{-10} 
\ \rm cm^{-2} s^{-1}$.

\begin{table}[htb]
\caption{Number of $\nu_\mu$ events with $E > 1$~TeV expected in 1 yr of observation.
We took into account that during propagation
 $\nu_\mu$'s will partition themselves equally between $\nu_\mu$'s and
 $\nu_\tau$'s due to maximal mixing~\cite{Learned:1994wg}.}

\vspace*{3mm}
\label{table}
\centering
\begin{tabular}{|c@{}|c|c@{}|c|c|}
\hline
\hline
~Experiment~ & \multicolumn{2}{@{}c}{ANTARES} & \multicolumn{2}{@{}c|}{NEMO} \\
\hline
\cline{1-3} \cline{4-5}
~$E_{\rm max}$ [TeV]~ & ~~$\Gamma=1.5$~~ & ~~$\Gamma=2.0$~~ & ~~$\Gamma=1.5$~~ & ~~$\Gamma = 2.0$~~ \\
\hline
\hline
10 & $0.20$ & $0.11 $  & 6.0   & 3.0  \\
\hline
100 & $0.26$ & $0.15$ & 8.3 & 5.0 \\
\hline
\hline
\end{tabular}
\end{table}

Future experiments in the Mediterranean Sea, such as
ANTARES~\cite{Aslanides:1999vq}, NESTOR~\cite{Aggouras:2005bg}, and
NEMO~\cite{Piattelli:2005mf} can provide meaningful probes of very 
high energy neutrinos from LS~5039. The most advanced project is ANTARES, with
construction beginning in the fall 2005 and completion expected by early
2007. This detector will have an instrumented area $> 0.06$~km$^2,$ with an
angular resolution better than $1.0^\circ$ ($0.2^\circ$) at energies larger than
100~GeV (100~TeV). In Table~\ref{table} we show event rates expected for
ANTARES assuming a power-law neutrino spectrum, ${\rm
d}N_{\nu_\mu}/{\rm d}E \propto E^{-\Gamma},$ with energy cutoff $E_{\rm max}=
10~{\rm TeV}$ and 100~TeV, and $\Gamma=$1.5 and 2. We used a realistic Monte Carlo simulation to account for the detector response. We also compensated for the fraction of the day in which the source is visible through upward-going muon tracks by ANTARES, which amounts to about 55\% duty cycle (for details see\cite{Montaruli:2004ze}). For all four combination of
parameters $\Gamma$ and $E_{\rm max}$ we assume the same energy flux of
neutrinos above 0.1 TeV, $F_{\rm E}=10^{-10} \ \rm erg/cm^2 s$. Note that this
is comparable with the energy flux corresponding to the dashed curve 1 in
Fig.~\ref{fig}, and a factor of 5 less than the upper bound on the $\nu_\mu$-flux 
set by COMPTEL data.  For a $1^{\circ}$ search cone, the atmospheric
neutrino background is at the level of 0.05 yr$^{-1}$; consequently in the most
optimistic scenario the full lifetime of the experiment will be required to
achieve a $5\sigma$ discovery. As we also show in Table~\ref{table}, more
promising data samples will be obtained by NEMO. The larger sensitivity and low
background ($\approx 1$~yr$^{-1}$) will allow NEMO to probe a broad  range of
plausible fluxes.

In summary, our  estimates of 
the neutrino flux associated with recent \textit{HESS}
discovery of TeV $\gamma$-rays from the direction of the microquasar 
LS~5039 show that the upcoming neutrino experiments
will be sensitive to this flux, provided that the observed 
TeV $\gamma$-rays are produced in the inner parts of the jet, and that the spectrum of 
the accelerated parent  protons extends to PeV energies.


\end{document}